\documentclass[twocolumn]{oja}
\usepackage{graphicx}
\usepackage{amsmath}
\usepackage{hyperref}
\usepackage{orcidlink}
\usepackage{lastpage}
\usepackage{xcolor}

\newcommand{\textsmall}[1]{\text{\footnotesize #1}}
\def\rsun{\ifmmode {\rm R_{\odot}}\else $\rm R_{\odot}$\fi}
\def\msun{\ifmmode {\rm M_{\odot}}\else $\rm M_{\odot}$\fi}
\def\mbh{\ifmmode {\rm M_{\bullet}}\else $\rm M_{\bullet}$\fi}
\def\AU{\ifmmode {{\text{au}}}\else au\fi}
\def\au{\AU}
\def\kms{\ifmmode {{\text{km s^{-1}}}}\else km s$^{-1}$\fi}
\def\yr{\ifmmode {{\text{yr^{-1}}}}\else yr$^{-1}$\fi}

\def\SgrAstar{Sgr~A$^\ast$}
\def\LMCstar{LMC$^\ast$}

\def\mbh{M_{\bullet}}
\def\mbin{m_{\rm bin}}
\def\abin{a_{\rm bin}}
\def\nbin{n_{\rm bin}}
\def\fbin{f_{\rm bin}}
\def\mavg{m_{\rm avg}}
\def\mstar{m_\ast}
\def\rstar{r_\ast}
\def\rtidal{r_{\rm tid}}
\def\abound{a_{\rm bnd}}

\def\rclose{r_{\rm close}}
\def\rD{r_{_{\rm D}}}
\def\rbhinflu{r_{v}}

\def\fgrav{f_\text{grav}}
\def\mbreak{M_b}
\def\mhills{M_{_{\rm Hills}}}

\def\vdisp{v_{\rm disp}}
\def\sigmabh{\sigma_\bullet}

\def\reptide{\texttt{REPTiDE}}

\begin{document}

\title{Supermassive black hole growth from stellar binary encounters}

\author{Aubrey L. Jones\orcidlink{0009-0008-6666-5896}}   % orcid.org
\author{Benjamin C. Bromley\orcidlink{0000-0001-7558-343X}}
\affiliation{Department of Physics and Astronomy, University of Utah, 115 S 1400 E, Salt Lake City, UT, 84112, USA}

\begin{abstract}
The growth of supermassive black holes (SMBHs) remains a central problem in astrophysics, with current observations providing only limited constraints on the underlying mechanisms. One possible growth channel is stellar accretion via the Hill's mechanism, wherein a SMBH tidally breaks up a passing binary star, capturing and eventually accreting a member of the binary. We adopt a framework based on kinematics to predict capture rates from parameters that include the central number density of stars, the stellar velocity dispersion, the binary fraction, and black hole mass. We then estimate the growth of SMBHs across a range of galactic environments. In a data set of 91 galaxies of various types and masses, we identify two candidates with SMBHs for which stellar accretion may be a driver of growth. Closer to home, a recent analysis of observed hypervelocity stars from the Large Magellanic Cloud (LMC) implicates binary star interactions with a massive black hole. Every hypervelocity star produced in this way leaves a bound partner that may be accreted, providing an active growth channel for the LMC's black hole.
\end{abstract}

\keywords{Supermassive black holes (1663) --- stellar dynamics (1596) --- tidal disruption (1696)}

\section{Introduction} 
\label{sec:intro}
Enshrouding millions to billions of solar masses within an event horizon of remarkably small size, supermassive black holes are famously elusive. Yet observations have teased out irrefutable evidence for their existence. In our own galaxy, years-long tracking of stars on tight orbits about a common point beautifully reveal \SgrAstar, the Milky Way's four million solar-mass black hole \citep{ghez2008, genzel2010}. The Event Horizon Telescope has since mapped gas flows near the supermassive black hole, unveiling a ``shadow'' surrounding the event horizon \citep{eht2022}.

Before these achievements were technologically feasible, J.~Hills (\citealt{hills1988}) sought other circumstantial evidence for supermassive black holes. Focusing on the Milky Way, Hills calculated that a binary star passing close to a SMBH in the center of the Galaxy would be broken up, with one stellar component flung out at high speed in a gravitational sling-shot, while its partner would remain bound to the black hole. The unbound star would be traveling fast enough to escape the Galaxy. Detection of a hypervelocity star (HVS) produced by this binary disruption (hereafter, ``break-up'') mechanism
would be compelling evidence of the black hole itself.

The discovery of a hypervelocity star by \citet{brown2005} confirmed Hills' prediction that a supermassive black hole in the center of the Milky Way produces fast stars with no other plausible origin. Subsequent detections \citep[e.g.,][see also \citealt{verberne2025}]{brown2006, brown2009, zheng2014} have built up a census of HVSs. The HVS population thus contributed to the growing body of evidence demonstrating that \SgrAstar\ was indeed a SMBH. More recently, based on Gaia astrometry \citep{gaiadr32023}, some of these same stars now appear to originate from the Large Magellanic Cloud (LMC). \citet{lmc} analyzed these sources to infer the existence of \LMCstar, a $\sim{6}\times 10^5$~\msun\ black hole in the dwarf galaxy. The Hills mechanism is fulfilling its promise as a signpost for SMBHs.

Hypervelocity stars are the lead astronomical actors in the Hills mechanism, yet they are only half the story. The bound partner in each binary break-up remains close to the black hole and eventually may be accreted onto it \citep[e.g.,][]{antonini2010, bromley2012, bradnick2017, yu2024, sersante2025}. A signature of this process is a tidal disruption event (TDE), which occurs when a black hole’s tidal forces shred the star \citep[see also \citealt{gezari2021} and references therein]{hills1975, frank1976, lightman1977, rees1988}. This outcome is possible for black holes with masses less than $\sim{10}^8$~\msun, for which the typical tidal disruption radius is outside the event horizon.  With more massive black holes, the star simply strays across the event horizon without a trace. 

Any bound star can be consumed, either tidally disrupted or swallowed by a black hole, once its orbital angular momentum drifts below a critical value, allowing it to plunge toward the event horizon \citep[e.g.,][]{merritt2013, correlations2023}. As fuel for SMBHs, higher-mass stars are more susceptible to tidal disruption because of their diffuse outer layers, which experience weaker gravitational binding forces. As a result, when these stars approach an SMBH, their outer envelopes are easily stripped away by tidal forces \citep{correlations2023}. These partial disruptions can occur multiple times before a massive star is completely destroyed \citep{compactsmbhb2022}. In any case, these stars, as well as their lower-mass counterparts, may ultimately be accreted by the SMBH \citep[e.g.,][]{thomsen2022, observation2022, olejak2025}. Stellar accretion may indeed be a dominant early-stage growth channel for black holes in the mass range of $10^{2}$--$10^6$~\msun, including single-star captures \citep{milosavljevi2006, stone2017} and binary break-ups \citep{gonzalezprieto2025}.

A variety of mechanisms have been proposed to explain the growth of supermassive black holes, including mergers and gas accretion \citep[e.g.,][]{silk1998, yu2002, dimatteo2003, montesinosarmijo2011, compactsmbhb2022}. Yet no single channel has emerged to account for the evolution of all SMBHs.  Here, our focus lies on the role of binary stars and the Hills mechanism as contributors to SMBH growth. We begin in \S\ref{sec:methods} by revisiting the theoretical description of this growth channel from \citet{bromley2012}, implementing it in a Python code to track binary star interaction rates with a SMBH and the subsequent evolution of the black hole mass. Next, in \S\ref{sec:christiandata}, we apply our analysis to galactic data provided by \citet[see also \citealt{hannah2025}]{hannah2024}, which span a wide range of black hole masses and galactic host properties. As a check on our methodology, we compare our approach to more sophisticated calculations of tidal disruption rates (\S\ref{sec:comparetde}). Closer to home, we assess the growth of \LMCstar, the black hole candidate in the Large Magellanic Cloud S\ref{sec:closertohome}. We summarize our results in \S\ref{sec:conclude}.

\section{The model: black hole growth through the Hills mechanism}
\label{sec:methods}

The Hills mechanism operates when the center of mass of a binary star comes within a distance $\rclose$ of a SMBH, where tidal forces from the black hole overcome the gravitational binding of the binary pair \citep{hills1988}. If $\rclose$ is within the event horizon of the SMBH (for black holes with mass $\mbh \gtrsim 10^8$~\msun), or inside the tidal radius of an individual star (for smaller black holes), the binary is certainly destroyed.  

The binary may be broken apart at larger closest-approach distances as well.  Following \citet[see also \citealt{yu2003}]{hills1988}, the closest-approach distance where disruptions are expected is 
\begin{eqnarray}\label{eq:rclose}
\rclose & \approx & \abin \left(\frac{\mbh}{\mbin}\right)^{1/3}
\\ \nonumber & \approx & 
260 \times
    \left[\frac{\abin}{1\ \textsmall{\au}}\right]
    \left[\frac{\mbh}{10^7~\textsmall{\msun}}\right]^{1/3}
    \left[\frac{\mbin}{0.6\ \textsmall{\msun}}\right]^{-1/3}  
    \text{\au.}
\end{eqnarray}
\citet{hills1988} noted that the break-up of a binary may occur at larger $\rclose$, depending on the binary's orbital phase and orientation. Conversely, some binaries may survive the encounter with the SMBH intact at smaller $\rclose$, again, depending on the details of their orbits.\footnote{\citet{hills1988} framed this range of outcomes probabilistically, with the likelihood of a break-up at $\rclose = 0$ being 100\%\ and decreasing linearly to zero at a closest-approach distance of $\rD/\abin$, where $\rD \approx \rclose (\mbin/11\mbh)^{1/3}$. Thus, when $\rD$ exceeds the binary separation (a value of about twice the distance in Eq.~(\ref{eq:rclose})), a binary can pass by a black hole unscathed.} Nonetheless, Equation~(\ref{eq:rclose}) provides a characteristic value of the binary break-up radius \citep{yu2003}. Below, in calculating interaction cross sections, we adopt an even smaller fiducial value of $\rclose = 200$~km/s, bringing the likelihood of a break up closer to unity.

We also consider the conditions in which a binary break-up would yield a bound partner within the sphere of influence of the black hole,
\begin{equation}\label{eq:rbhinflu}
\rbhinflu = \frac{G \mbh}{\vdisp^2} 
 \approx 4.3 \times 
    \left[\frac{\mbh}{10^7~\textsmall{\msun}}\right]
    \left[\frac{\vdisp}{100~\textsmall{km/s}}\right]^{-2}
    \text{pc.}
\end{equation}
The semimajor axis of the bound partner after tidal break-up is 
\begin{eqnarray}\label{eq:abound}
\abound & = &  0.055 \times 
    \left[\frac{\abin}{1~\textsmall{\au}}\right]
    \left[\frac{m_\text{ej}}{0.3~\textsmall{\msun}}\right]^{-1}  \\ \nonumber
    & \ & 
    \times \left[\frac{\mbin}{0.6~\textsmall{\msun}}\right]^{1/3}    
    \left[\frac{\mbh}{10^7~\textsmall{\msun}}\right]^{-1/3}
    \text{pc}
\end{eqnarray}
where $m_\text{ej}$ is the mass of the ejected partner. This expression quantifies the limitations of binary disruption for wide binaries. In the case of a $10^7$~\msun\ SMBH, the bound partner of a binary with stars of mass 0.3~\msun\ would end up outside of the black hole's sphere of influence if $\abin \gtrsim 80$~\au. If a similar binary encountered a $10^5$~\msun\ black hole, neither star would be captured within the sphere of influence if the binary separation were much larger than 1~\au.

\subsection{Binary break-up rates}

Here we are interested in representative binary properties to predict the change in mass of a SMBH from the accretion of stars captured by the Hills mechanism. Our starting point for the rate is ``$nv\sigma$'' from kinematics, where $n$ is the number density of binaries, $v$ is their mean speed in a region surrounding a SMBH, and $\sigma$ is the cross-sectional area of the black hole --- the size of the hypothetical target for a binary star to hit in order to be split into a hypervelocity star and a bound partner. We consider each factor in turn.

We express the number density of binary stars in the vicinity of a SMBH, $\nbin$, in terms of the central stellar mass density of its galactic host, $\rho_0$, treated as a constant. If the fraction of binaries in this region is $\fbin$, then 
\begin{equation}
\nbin \approx \frac{\rho_0}{\mavg}\frac{\fbin}{(1+\fbin)},  
\end{equation} 
where $\mavg \approx$ 0.3 \msun\ is the average mass of a star, counting binary components as individual stars. In the Milky Way, the binary fraction is roughly half of all stars \citep[and references therein]{duchene2013}, although $\fbin$ is dependent on the spectral type of the components \citep[e.g.,][]{moe2017}. Typically younger, more massive stars have a higher binary fraction than less massive, older stars. 

The central mass density of stars, $\rho_0$, strongly depends on the SMBH mass, $\mbh$. For large elliptical galaxies, the density falls with increasing black hole mass $\rho_0 \sim 1/\mbh$ \citep[e.g.,][]{faber1997}. The trend is opposite in smaller galaxies. Both central stellar density and black hole mass increase with the size and mass of the galactic host. Here, we introduce the parameterization,
 \begin{equation}\label{eq:massdens}
    \rho_0 
     = 5\times 10^4 \frac{\mbh/\mbreak}{1 + \mbh^2/\mbreak^2}
    % = 5\times 10^4 \frac{\mbh \mbreak}{\mbh^2+ \mbreak^2}
    \ \ \text{\msun/pc$^3$},   
\end{equation}
where $\rho_0$ is the central density within some small (parsec-scale) distance of the SMBH, and $\mbreak \approx 4\times 10^7$~\msun\  is the black hole mass associated with the break between two power-law relationships; $\rho_0 \sim \mbh$ in the low-mass regime and $\rho_0 \sim 1/\mbh$ for masses much greater than $\mbreak$. 

Equation~(\ref{eq:massdens}) is intended to provide representative values of the stellar mass density. Our parameterization in that equation was inspired by galaxy data kindly provided by \citet{hannah2024}, wherein the stellar mass density is evaluated at 5~pc away from the central SMBH. We demonstrate the effectiveness of Equation~(\ref{eq:massdens}) in the next section. For a more detailed description of the stellar mass density in the centers of galaxies, we assume a power-law profile in radius $r$, so that  
\begin{equation}\label{eq:densprofile}
    \rho(r) = \rho_0 \left(\frac{r}{r_0}\right)^{\gamma},
\end{equation}
where $\gamma \approx -2$ \citep{bahcall1976} and we set $r_0 = 5$~pc. We make use of this expression in the next section as well.

The central velocity dispersion of stars in the cores of galaxies, $\vdisp$, is typically of the order of 100~km/s, with values that increase with SMBH mass and total galactic mass \citep[e.g.,][]{faber1997, bernardi2003, mcconnel2013, greene2020}. The range of observed values is narrow compared with that of the masses of known SMBHs, because 
the dispersion has weak dependence on black hole mass:
\begin{equation}\label{eq:vdisp}
    \vdisp \propto \mbh^{1/4}
\end{equation}
\citep{kormendy2013} and broader galactic properties including luminosity \citep[e.g.,][]{faber1976}, which can span many orders of magnitude. 

The third factor in the kinematic description is $\sigmabh$, the cross-sectional area of the black hole in scattering events with binaries. This term depends on uncertain quantities including the distribution of binary orbital elements, and the nature of their orbits around the black hole. Following our review of the Hill mechanism above, we adopt $\rclose$ (Eq.~(\ref{eq:rclose})) for calculating the geometric cross-section near the black hole ($\pi \rclose^2$). Gravitational focusing by the SMBH boosts the geometric cross section by a factor of  
\begin{equation}\label{eq:fgrav}
\fgrav \approx 1 + \frac{2G\mbh}{\rclose\vdisp^2},
\end{equation}
where the second term dominates for stellar orbits around supermassive black holes \citep{bondi1944}. The aim of a binary toward the black hole may be wide, yet gravity draws it in. 

Together, the three ``$nv\sigma$'' factors provide an estimate of the rate of binary break-up through the Hills mechanism. In this kinematical approximation, the break-up rate of binary stars through the Hills mechanism is
\begin{eqnarray}\label{eq:k}
    k & = & \frac{\rho_0}{\mavg}\frac{\fbin}{(1+\fbin)} \cdot \vdisp \cdot \pi \fgrav \rclose^2
    \\ \nonumber
    \label{eq:kfid}
    & \approx & 0.008 % 5.5\times 10^{-3} % 1.1\times 10^{-3}
    \left[\frac{\rho_0}{10^4\ \textsmall{\msun/pc}^3}\right]
    \left[\frac{\mavg}{0.3\ \textsmall{\msun}}\right]^{-1}\!
    \left[\frac{\fbin}{0.1}\right]
    % \times 
    \\ \nonumber
    & \ & \times
    \left[\frac{\vdisp}{100~\textsmall{km/s}}\right]^{-1}
    \left[\frac{\mbh}{10^7~\textsmall{\msun}}\right]
    \left[\frac{\rclose}{200~\textsmall{\au}}\right] \ \text{yr}^{-1}; 
    % M^(rho_index + 0.25 + 1 -1/3 = (i low mass) 2 - 1/12 = 23/12 (ii high mass) 13/12 
\end{eqnarray}
in the lower expression, we have simplified the dependence on $\fbin$ (formally valid for $\fbin \ll 1$) and the gravitational focusing factor (since $\fgrav \sim \mbh$ when $\fgrav \gg 1$, which holds for all SMBHs and stars in their proximity). This value represents an upper bound, as the fiducial central density $\rho_0$ is near its maximum in Equation~(\ref{eq:massdens}). 

By considering the explicit SMBH mass dependence of the central density (Eq.~(\ref{eq:massdens})), the trend between black hole mass and velocity dispersion (Eq.~(\ref{eq:vdisp})), and the interaction cross-section with gravitational focusing ($\fgrav\rclose^2 \propto \mbh^{5/6}$ \citep{hills1988}, we estimate how the SMBH mass affects the binary break-up rate: 
\begin{equation}\label{eq:kmbh}
    k \propto \frac{\mbh^{25/12}}{\mbreak^2 + \mbh^2} 
 \rightarrow 
\begin{cases}
\mbh^{2.08} & \text{if } \mbh \ll \mbreak, \\
\mbh^{0.08} & \text{if } \mbh \gg \mbreak.
\end{cases}
\end{equation}
%These expressions suggest 
This equation suggests  a ``sweet spot'' of $\mbh\approx \mbreak$, around  $10^7$~\msun\ where the fractional rate $k/\mbh$ reaches a peak. Next we explore how these results translate into black hole growth.

\subsection{A channel for black hole growth}\label{subsec:channel}

Every interaction between a binary star and a supermassive black hole that leads to binary break-up places a star on a bound orbit that will eventually be captured by the SMBH as a result of orbital diffusion \citep[e.g.,][]{bromley2012}. Although tidal disruption outside the horizon may not result in accretion of all the freed stellar gas, at least initially \citep{evans1989}, we assume that for stars already on bound orbits, all the disrupted gas ultimately is accreted by the SMBH. In this scenario, the binary break-up rate in Equation~(\ref{eq:k}) directly provides the black hole growth rate, 
\begin{equation}\label{eq:dmdteqk}
    \frac{d\mbh}{dt} = k \mavg,
\end{equation}
from accretion of captured Hills' stars.  To solve this equation, we make the following definitions and assumptions:
\begin{itemize}
    \item The black hole mass $\mbh$ is a function of time $t$, with an initial mass $\mbh(0)$ at $t=0$, the present epoch. 
    \item The central stellar density $\rho_0$ is also a function of time through its dependence on $\mbh(t)$ in Equation~(\ref{eq:massdens}), scaled so that $\rho_0(0)$ is the present-day value.
    \item The gravitational focusing factor $\fgrav$ depends on $\mbh(t)$, with $\fgrav \sim \mbh$, since $\fgrav \gg 1$  for SMBHs in the cores of galaxies (see Eq.~(\ref{eq:fgrav})).
    \item All other parameters ($\rclose$, $\vdisp$, $\fbin$) are set to representative constant values, since taken together, they have weak dependence on the SMBH mass. 
\end{itemize}

With this prescription in the growth equation (Eq.~(\ref{eq:dmdteqk})), we obtain an ordinary differential equation of the form
\begin{equation}\label{eq:dmdt}
    \frac{d\mbh}{dt} = A \frac{\mbh^2}{\mbh^2+\mbreak^2},
\end{equation}
where $A = k\mavg$ is a constant (cf.~Eq.~(\ref{eq:kfid})). Its solution, from separation of variables, is
\begin{equation}\label{eq:msolve}
    \mbh(t) = \frac{1}{2}\left(X + \sqrt{X^2+4\mbreak^2}\right)
\end{equation}
where $X = At + \mbh(0) - \mbreak^2/\mbh(0)$ and $\mbh(0)$ is the black hole mass at $t=0$, the present epoch. This approximation gives a good representation of growth in most cases of interest here, wherein the SMBH mass increases by some fractional amount over the course of a gigayear.

\begin{figure}
    \centering
    \includegraphics[width=1\linewidth]{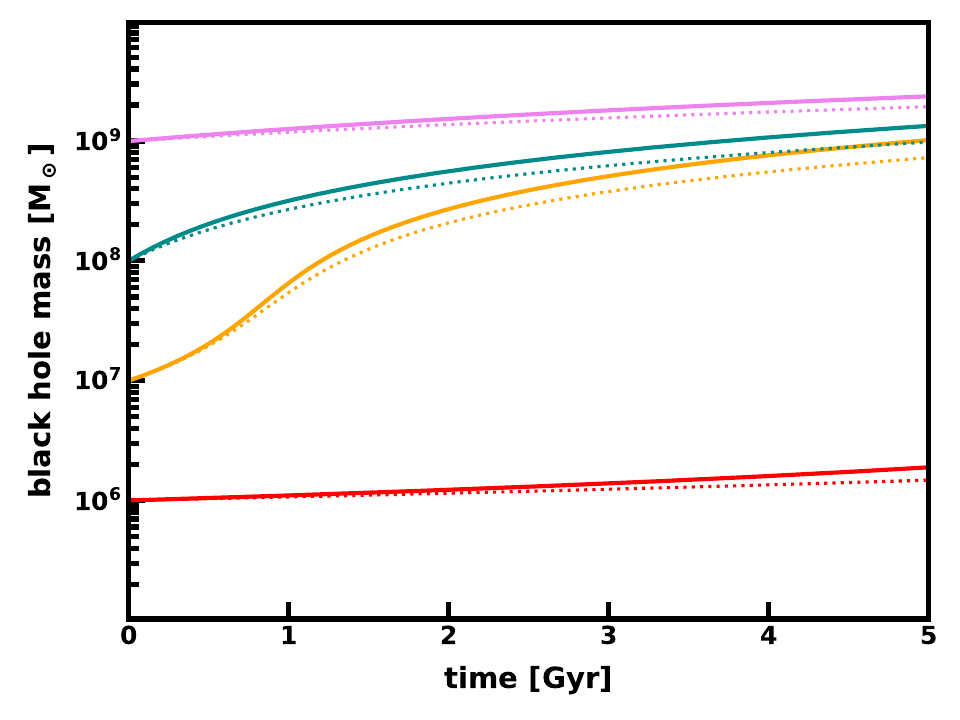}
    \caption{The evolution of black hole mass from the Hills mechanism and stellar accretion. Each curve shows a unique starting mass, which we numerically evolve using the growth equation (Eq.~(\ref{eq:dmdt})). We adopt the analytical relationship between black hole mass and central stellar density in Equation~(\ref{eq:massdens}), and include the mass-dependence of other parameters based on the fiducial values in Equation~(\ref{eq:kfid}). The analytical solutions (Eq.~(\ref{eq:msolve}), dashed curves) are shown for comparison. Because the central density is near a maximum for $\mbh \approx 4\times 10^7$~\msun, the most rapid growth occurs for black holes around this size.
    \label{fig:smbhgrowlog}}
\end{figure}

We also use a step-wise numerical integrator (the Euler method), written in Python, to solve for black hole growth including the variation of all parameters with SMBH mass (Eq.~\ref{eq:kmbh}). We start with the initial SMBH mass $\mbh(0)$ and calculate $k\mavg$. The mass at a time $t = \Delta t$ later is $\mbh(0) + k\mavg \Delta t$. This process is repeated until we reach some desired end time using $\Delta t = 0.1$--1~Myr. This numerical approach allows us to explore more complicated parameterizations and growth scenarios.  Figure~\ref{fig:smbhgrowlog} provides an illustration.

\subsection{A modified kinematic model}

The rapid growth of a SMBH with masses greater than $\mbreak \approx 10^8$~\msun\ predicted by the simple kinematical model in Equation~(\ref{eq:dmdt}) may be challenging to realize. Growth may be limited by ``loss-cone'' effects, wherein binary break-ups and the accretion of stars generally proceed at a much slower rate than predicted by simple kinematics by virtue of the depletion of orbits that pass close to the SMBH \citep[see also \citealt{hannah2025} and references therein]{frank1976, lightman1977, cohn1978, magorrian1999, perets2007, merritt2013}. Furthermore, even if phase-space mixing rapidly replenished orbits in the loss cone, the mass reservoir in stars surrounding the black hole may insufficient to affect its overall mass. 

For a better assessment of growth rates in light of loss-cone effects and limited mass reservoirs surrounding SMBHs, we introduce two changes. First, we modify the stellar mass density in the rate estimation to the value not at a fixed distance (5 pc) from a SMBH, but at the outer limit of the black hole's sphere of influence to represent unbound stars (Eq.~(\ref{eq:rbhinflu})). Then, to enforce the depletion of orbits in loss cones around the most massive SMBHs, we introduce a factor of $1/(1+\mbh/\mbreak)$. The result is an ansatz, a modified kinematical model,
\begin{equation}\label{eq:dmdtmod}
    \frac{d\mbh}{dt} = A \mbreak\rbhinflu^{-\gamma} \frac{\mbh^{2+\gamma}}{(\mbh^2+\mbreak^2)^2},
\end{equation}
where $A$ is the same constant as in Equation~\ref{eq:dmdt} and $\gamma$ is the power-law index for the stellar density profile (Eq.~(\ref{eq:densprofile})). We validate this new parameterization by comparing with detailed TDE rates in \S\ref{sec:comparetde}. We next explore black hole growth in observed galaxies with both the kinematical model and this modified version.

\section{SMBH growth in nearby galactic hosts}
\label{sec:christiandata}

We apply our binary break-up models to observed galactic sources to evaluate the potential growth of supermassive black holes from binary break-up and stellar accretion under varying conditions. While many of the model parameters are unknown, estimates are available for the two critical ones, then SMBH mass $\mbh$ and the central stellar density $\rho_0$. Here, we consider black hole growth using the curated dataset from \citet{hannah2024},  consisting of 91 galaxies of a range of types, all within 50~Mpc. With the Python Pandas package, we highlight data by filtering to include black holes with mass derivation methods corresponding to source codes 1, 2, and 3, ensuring methodological consistency. We thus identify 30 well-measured galaxies to showcase in this study. The remaining 61 sources have black hole masses that are not as well constrained. For each selected galaxy, we extracted the galaxy name, the black hole mass (in logarithmic solar mass units) and central stellar density (in logarithmic solar masses per cubic parsec), defined as the value at a distance of 5~pc from the dynamical center of the galaxy.

\begin{figure}
    \centering
    \includegraphics[width=1\linewidth]{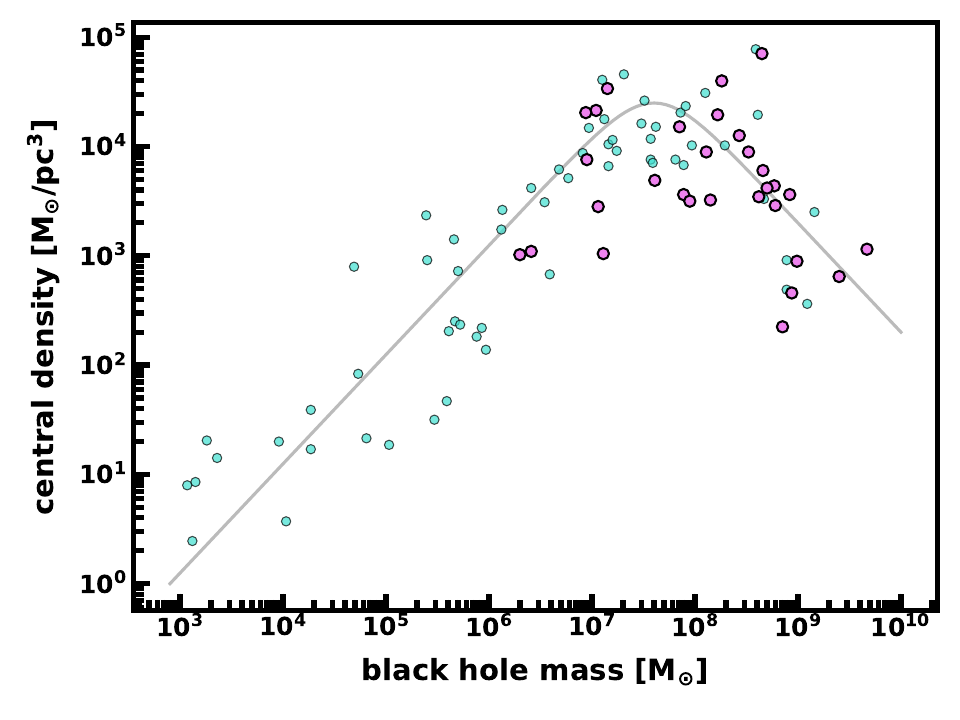}
    \caption{Central density versus SMBH mass. The larger, violet points correspond to sources with more reliable black hole mass estimates, while the smaller, turquoise points represent galaxies with poorly constrained black hole masses. The gray line is a simple analytical model for the mass-density relation (Eq.~(\ref{eq:massdens})). The low-mass end of this model is speculative. In all cases, the stellar densities are evaluated at 5~pc from the galactic centers.}
    \label{fig:massdens}
\end{figure}

Figure~\ref{fig:massdens} illustrates the dataset, showing the representative central stellar mass density (evaluated at 5~pc from the galactic center) versus SMBH mass for all sources in \citet[see also \citealt{hannah2025}]{hannah2024}. The plot also shows our analytical expression that broadly tracks the central density as a function of black hole mass (Eq.~(\ref{eq:massdens})).

\begin{figure*} % * causes colspan = 2 in two column mode...
    \centering
    \includegraphics[width=0.85\linewidth]{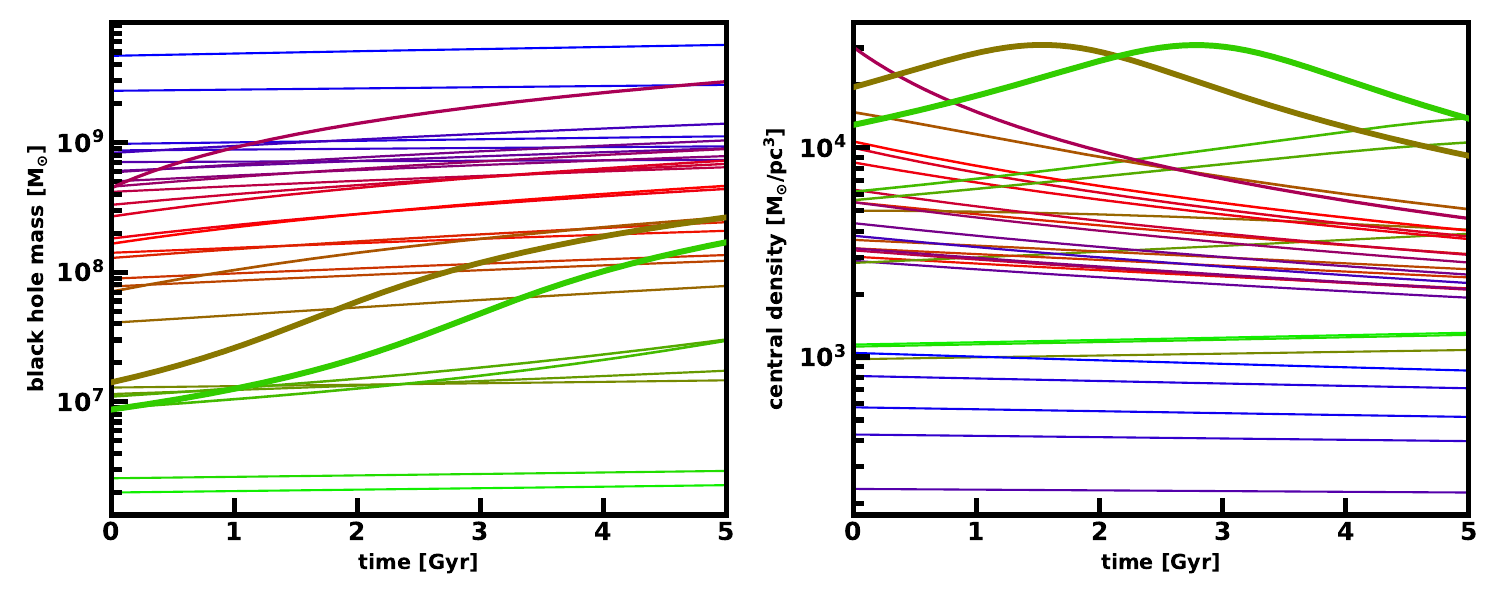} 
    \caption{Black hole mass and central density evolution from the Hills mechanism growth channel predicted for the 30 best-measured galaxies with SMBHs in the \citet{hannah2024} sample. We begin with the observed masses and central densities, the use these values to predict future growth of the SMBH, as described in the main text (\S\ref{subsec:channel}). The colors run from green hues (low-mass SMBHs) through reds (intermediate massed) to blue (high-mass SMBHs), and the same unique color is assigned to each galaxy. The thickness of the lines correlates with the relative increase in black hole mass over the 5~Gyr run. Clearly, growth favors the black hole in a high-density field of stars.
    \label{fig:smbhgrowdata}}
\end{figure*}

To assess the potential for growth, we start with the estimated SMBH mass of each galaxy and the reported central stellar mass density. We then integrate the mass forward in time as in Figure~\ref{fig:smbhgrowlog} using our kinematical model (Eq.~(\ref{eq:dmdt})). The central stellar density $\rho_0$ is given the same form of dependence on black hole mass as in Equation~(\ref{eq:massdens}), but the normalization is set to match the density in \citet{hannah2025}. Our results are presented in Figure~\ref{fig:smbhgrowdata}, which shows predicted future growth of the 30 galaxies from the \citet{hannah2024} dataset with the most reliable SMBH masses.

\begin{figure}
    \centering
    \includegraphics[width=1\linewidth]{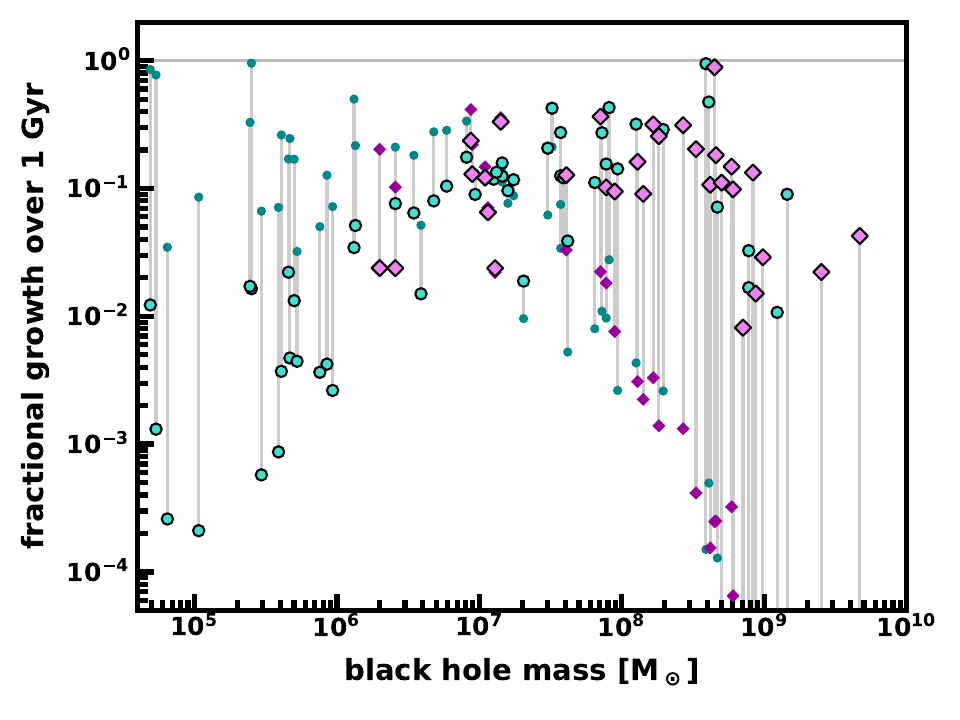}
    \caption{Fraction of the black hole mass predicted to have come from stellar accretion through the Hills mechanism. The diamond symbols are the predictions for the 30~galaxies with high-quality SMBH masses in the \citet{hannah2025} data set, run backwards in time for 1~Gyr with our numerical integrator. 
    \label{fig:smbhgrowbacktrace}}
\end{figure}

Another view of black hole growth comes from integrating the growth equation from the simple kinematical (``$nv\sigma$'') model (Eq.~(\ref{eq:dmdt})) backward in time. In this way, we predict the fraction of a black hole's mass that could potentially have come from the Hills mechanism and stellar accretion. Figure~\ref{fig:smbhgrowbacktrace} provides that view. It also shows the predictions from the modified model with stellar density set to its value at the radius of influence of the black hole and a term to suppress growth of black holes with $\mbh \gtrsim \mbreak$ (Eq.~(\ref{eq:dmdtmod})).

The predicted mass evolution in Figures~\ref{fig:smbhgrowdata} and \ref{fig:smbhgrowbacktrace} illustrates that conditions for black hole growth vary widely among galaxies and our model assumptions. Of the 91 sources in the \citet{hannah2024} data set, the SMBHs in 39 galaxies are predicted to have grown in mass by more than 10\%\ in the last gigayear, according to the simple kinematical model. All of these galaxies have SMBHs with masses near or above $10^7$~\msun. In the modified model, 37 galaxies have SMBHs that grew by that same amount in a gigayear, and most of these galaxies have lower-mass SMBHs. A total of 11 galaxies have SMBHs that have grown by 10\%\ or more in both models. Table~\ref{tab:candidates} lists these galaxies.

\begin{deluxetable}{lccccc}
\tabletypesize{\footnotesize}
\tablecaption{Promising candidates for black hole growth by binary break-up
\label{tab:candidates}}
\tablehead{
   \colhead{\ } & 
   \colhead{$\log_{10}(\mbh)$} & 
   \colhead{$\log_{10}(\rho_0)$} &
   \multicolumn{2}{c}{growth [\%]} &
   \colhead{qual.} 
   \\
   \colhead{name} & 
   \colhead{\scriptsize [\msun]} & 
   \colhead{\scriptsize [\msun/pc$^3$]} &
   \colhead{$nv\sigma$} & 
   \colhead{mod.} & 
   \colhead{flag}}
\startdata 
 NGC 2778 &   7.150 &   4.290 &   33.28 &   35.76 & 1 \\
 NGC 3384 &   7.040 &   3.750 &   12.16 &   14.78 & 1 \\
 NGC 3522 &   7.100 &   3.730 &   11.82 &   12.36 & 0 \\
 NGC 3945 &   6.940 &   4.110 &   23.66 &   41.50 & 1 \\
 NGC 4387 &   7.160 &   3.750 &   12.48 &   11.24 & 0 \\
 NGC 4458 &   7.120 &   3.790 &   13.41 &   12.70 & 0 \\
 NGC 4483 &   7.160 &   3.870 &   15.85 &   14.60 & 0 \\
 NGC 4612 &   6.770 &   3.700 &   10.44 &   28.54 & 0 \\
 NGC 4638 &   7.510 &   4.370 &   42.50 &   21.11 & 0 \\
 NGC 5236 &   6.910 &   3.950 &   17.57 &   33.67 & 0 \\
 NGC 7457 &   6.950 &   3.790 &   12.95 &   22.00 & 1 
 \enddata
\tablecomments{Columns 4 and 5 are the percent growth of the SMBH over the past gigayear, assuming the simple kinematic model (column 4, ``$nv\sigma$'', Eq.~(\ref{eq:dmdt})) and the modified version  (column 5, ``mod.'', Eq.~(\ref{eq:dmdtmod})). The last column on the right indicates the reliability of the SMBH mass measurement from \citet{hannah2024}. High-quality mass estimates (source codes 1, 2, and 3 in \citealt{hannah2024}) are indicated with a ``1'' while less reliable sources have a ``0'' designation in that column.}
\end{deluxetable}

While a few lower-mass SMBHs experience rapid growth (e.g., SMBHs in NGC 4605 and NGC 5238, with growth of over 90\%\ in the past gigayear) in the modified kinematic model, and a couple of high-mass SMBHs (in NGC 4660 and NGC 4342) similarly stand out in the simple model, only the 11~galaxies in Table~\ref{tab:candidates} have significant SMBH growth in both models. Of these, two --- NGC 2778 and NGC 3945 --- are standouts because they have reliable SMBH mass measurements ($\mbh \sim 10^7$~\msun) and SMBH growth rates above 20\%\ per gigayear. 

NGC 2778 is an elliptical galaxy with a central velocity dispersion of $\vdisp \approx 150$~km/s \citep{jin2025} and a mass $\mbh = 1.4\times 10^7$~\msun. The growth rate in this case may be overstated because the stellar velocity dispersion is higher than our adopted value by a factor of about 40\%. Given that the binary fraction we have adopted is modest ($\fbin = 0.1$), we still expect that this galaxy may be a promising candidate for the binary break-up growth channel.

The galaxy NGC 3945 has a barred lenticular morphology with prominent rings and a central velocity dispersion of about 170~km/s \citep{gultekin2009}. While this dispersion is also high compared with our prescription given $\mbh \approx 10^7$~\msun,  it still leads to a predicted growth rate of 13--24\% per gigayear, depending on the kinematic model. Furthermore, the non-axisymmetric distribution of stars in this barred galaxy offers promise of phase-space mixing and perhaps loss-cone replenishment beyond the SMBH's sphere of influence \citep[e.g.,][]{sellwood1993}. 

We conclude that for many of the galaxies considered here, the binary-capture channel may play a significant, and, in some cases, even a dominant role in black hole growth.

\section{Comparison with predictions of tidal disruption events}\label{sec:comparetde}

Our estimates of SMBH growth from binary break-up are based on encounter rates from simple ``$nv\sigma$'' kinematics, a choice that highlights what might be possible, not necessarily what is plausible. \citet{hannah2025} performed more detailed calculations for stellar encounters that lead to TDEs for the same set of galaxies considered here. They used a sophisticated phase-space evolution code, \reptide\ \citep{reptide}, with inputs that include the stellar mass density profile (density at 5~pc and the power-law index $\gamma$) and black hole masses. Their published TDE rates offer an opportunity for us to connect the impact of assumptions made here for binary break-up.

We compare our approach with \reptide\ results by adapting our simple and modified kinetic models to stellar disruption instead of binary break-up. In the event rate estimate (Eq.~(\ref{eq:k})), we remove dependence on $\fbin$ and replace $\rclose$ with a representative stellar tidal radius 
\begin{equation}
    \rtidal = \rstar \left(\frac{\mstar}{\mbh}\right)^{1/3}
\end{equation}
where $\mstar$ is the star's mass (here, $\mstar = \mavg = 0.3$~\msun) and $\rstar \propto \mstar^{0.8}$ is its radius. 

\begin{figure}
    \centering
    \includegraphics[width=1\linewidth]{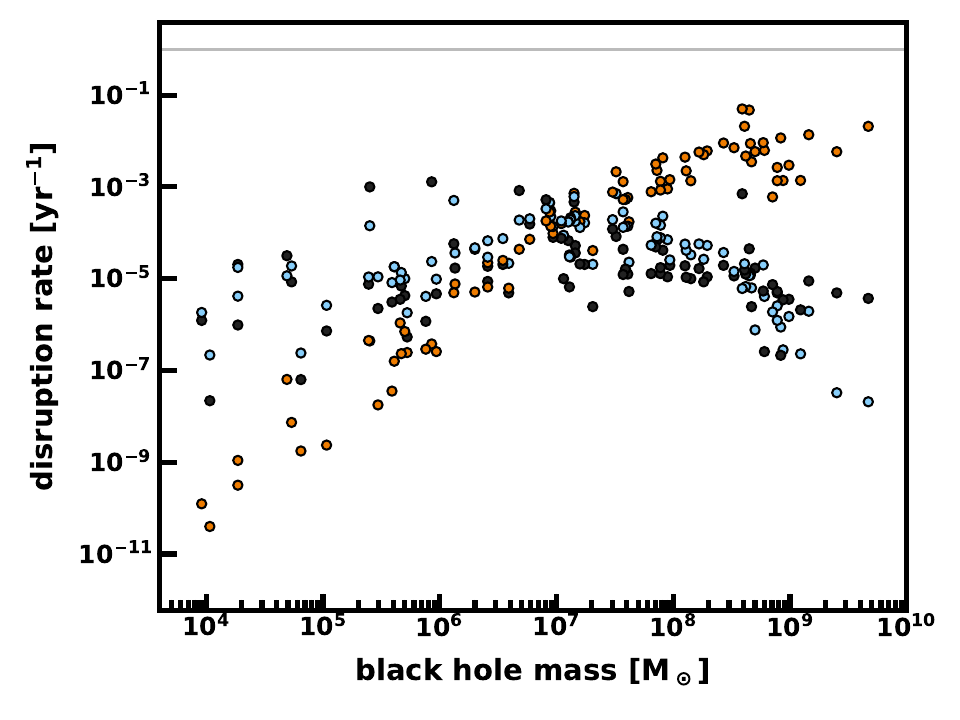}
    \caption{Tidal disruption event rates from \reptide\ \citep{hannah2025} and the two kinematic models presented here, adapted to encounters between single stars and a SMBH. The black symbols are the predictions from \reptide, while the dark orange symbols are from our simple kinematic model. Our model overestimates the TDE rate at large SMBH mass, and underestimates it at low mass. The modified kinematic model predictions (light blue symbols) better track the \reptide\ estimates.}
    \label{fig:compare_tde}
\end{figure}

Figure~\ref{fig:compare_tde} is the result. The trend in the \reptide\ predictions is broadly decreasing TDE rates with SMBH mass. \citet{hannah2025} note that the stellar phase-space distribution for galaxies with smaller SMBH masses have ``full loss cones''; there is little depletion of radial orbits that can take stars close to the SMBH. For higher-mass black holes, the loss-cone becomes depleted. In our simple kinematic model, we assume a full loss cone in all cases. Thus, our predictions of TDE rates for the highest-mass black holes are relatively high. For lower-mass SMBHs, our model predicts a comparatively low rate. The reason is that the stellar density in our simple model is assessed at 5~pc, well beyond the SMBH radius of influence, and likely an underestimate of the density of unbound stars that interact with the SMBH.

In contrast, our modified kinematical model, which uses an estimate of the stellar mass density just outside the sphere of influence, broadly matches the \reptide\ results for SMBH masses up to $\mbh \sim 10^9$~\msun. Indeed, our parameterization in Equation~(\ref{eq:dmdtmod}) was guided by this comparison.

The success of our modified kinematic model in reproducing TDE rates from loss-cone theory calculations highlights a liability. Binary break-up rates can be significantly higher than loss-cone predictions because the interaction cross section depends on the orbital separation, which is much larger than the tidal radius of main-sequence stars. Yet even loss-cone theory for the capture of single stars may over-predict observed TDE rates. For example, \citet{yao2023} analyzed TDEs observed in optical light and concluded that TDE event rates per galaxy were well below $10^{-4}$ per year. Low-number statistics, waveband selection, and flux limits \citep[\S4 therein]{hannah2025}, as well as accounting for partial TDEs \citep[e.g.,][]{bortolas2023}, may all play a role in resolving this tension between theory and observation. However, an increase in the predicted rate of TDEs from captured binary partners on tight orbits may be challenging.

We offer the following considerations that might support the role of binary break-up in SMBH growth. Black holes with mass greater than the Hills mass,
\begin{equation}
    \mhills = \left(\frac{c^2 \rstar}{2 G \mstar^{1/3}}\right)^{3/2},
\end{equation}
will accrete solar-type stars without an observable TDE, since tidal disruption occurs inside the horizon \citep{hills1975}. For stars of mass $\mavg$, $\mhills$ turns out to be close to $\mbreak$. If the predictions of the simple kinematic model (binary star-SMBH encounter rates increase with black hole mass) have relevance to these larger black holes, then binary break-up is an avenue for depositing stars deep inside a SMBH's sphere of influence (Eqs.~(\ref{eq:rbhinflu}) and (\ref{eq:abound})) irrespective of observed constraints on TDE rates. Evidence for this potential growth channel would necessarily be indirect, possibly from a gravity wave signature \citep[e.g.,][]{toscani2025}.  

Lower-mass black holes with $\mbh \lesssim 10^6$~\msun\ (including intermediate-mass black holes) can experience significant growth even with the low TDE rates inferred from observations \citep{milosavljevi2006, stone2016}. Binary break-up may be a promising avenue to bring stars close enough a lower-mass central black hole to be fully disrupted and accreted. We explore a compelling example, next, wherein binary break-up is directly implicated.    

\section{Closer to home: the case of LMC*}
\label{sec:closertohome}

\citet{lmc} recently proposed the existence of a $6\times 10^5$~\msun\ black hole within the Large Magellanic Cloud (LMC). Their inference is based on the locations and velocities of the observed HVSs from the HVS Survey \citep{brown2012}. This analysis incorporated Gaia Data Release 3 astrometry \citep{gaiadr32023} to reveal that half of the HVS Survey sources originated in the LMC. With forward modeling using the parameterization of the ejection speed of \citet{bromley2006} and an HVSs production rate of $k = 2$~Myr$^{-1}$, \citet{lmc} reproduced the observed phase-space distribution of these HVS, including the velocity increase from LMC motion in the Galactic frame and the observed clustering in the sky within the context of the survey selection. The success of this modeling requires the presence of \LMCstar.

The results in \citet{lmc} have implications for the growth of the dwarf galaxy's massive black hole. The observed cluster of nine main-sequence, B-type HVSs linked to an LMC origin, found in the constellation Leo, turn out to be the tip of an iceberg representing binary star encounters with \LMCstar. First, these stars are at the edge of the HVS Survey window, which do not include a large swath of sky around the LMC where the majority of HVSs are expected to reside. From the reported rate of HVS production required to match observations, these nine stars are associated with $\sim$800 other ejected stars with similar masses that lie beyond the surveyed sky. Second, B-stars are comparatively rare, and are the tip of another iceberg from a stellar population perspective. With a Salpeter initial mass function (IMF), stars with mass $m \geq 2$~\msun\ --- the range considered by \citet{lmc} --- represent less than 2\%\ of all stars with masses greater than 0.1~\msun. The nine HVSs from \LMCstar\ therefore suggest a binary break-up rate greater than $10^{-4}$~yr$^{-1}$. With an average mass around $0.3$~\msun\ captured on tight orbit around \LMCstar\ per break-up, the SMBH can acquire a significant fraction of its mass over a few gigayears.

There are important uncertainties in this growth rate. A major source is the initial mass function, since the HVS Survey only sampled the most massive stars. A top-heavy IMF would reduce the rate we infer here --- the observed stars then are the tip of a less significant iceberg. Other sources of uncertainty include details that affect the Hills mechanism and its outcome, including the distribution of binary orbits, the periastron distribution \citep[empty or full loss cone][]{lightman1977, stone2016}, and the mass distribution around the SMBH that impacts the speed (and interpretation) of the HVS candidates \citep[e.g.,][]{gould2003, bromley2006, sari2010, rossi2014, lmc}.

A check on our overall approach (\S\ref{subsec:channel}) is a comparison between our predicted binary break-up rate and the \citet{lmc} LMC HVS study. We adopt a velocity dispersion of 35~km/s, characteristic of the central region of the LMC \citep{vasiliev2018}, 
%(in absence of a compelling value from the literature) 
a central density at 5~pc of $\rho_0 \approx 750$~\msun/pc$^3$ from Equation~(\ref{eq:massdens}), and our fiducial binary fraction of $\fbin = 0.1$ \citep[likely an underestimate, cf.][]{dalton1995}. Absent a measurement of a stellar nucleus, we use $\gamma = -7/4$ to extrapolate $\rho$ inward from 5~pc (Eq.~(\ref{eq:densprofile})) in our modified kinematic model. Our predicted binary break-up rates in the simple and modified models are $k = 4\times 10^{-5}$~yr$^{-1}$ (simple) to $k = 2\times 10^{-4}$~yr$^{-1}$ (modified), in the range of an estimate based on the HVS number count reported in \citet{lmc}.  At these rates, between 10\%\ and 40\%\ of \LMCstar's mass came from stellar accretion over the past five gigayears. A modest increase in the assumed binary fraction would correspond to an even greater role of binary break-up for the black hole's growth. 

Although binary accretion on its own cannot account for the emergence of a $10^6$~\msun\ black hole, the mechanism can help build a SMBH from a smaller seed object. A similar result was found in a recent detailed study of black hole growth in globular clusters \citep{gonzalezprieto2025}. 

\section{Conclusion}\label{sec:conclude}
 
We present a simple theoretical framework to assess SMBH growth from accretion of stars captured through binary break-up (\S\ref{sec:methods}). Our approach is based on kinematical rates, including a simple ``$nv\sigma$'' model and a modified version designed to accommodate loss-cone effects and limited reservoirs of stellar mass. We apply our models to a set of nearby galaxies hosting SMBHs, predicting both past and future growth by this mechanism (\S\ref{sec:christiandata}). A comparison between tidal disruption event rates and our approach (\S\ref{sec:comparetde}) suggests that our modified kinematical model realistically tracks the encounter rate of binary stars, providing some confidence in our predictions. Finally, in \S\ref{sec:closertohome}, we examine black hole growth in the LMC, following the recent analysis of \citet{lmc} and observed hypervelocity stars that directly reveal binary break-ups.

Our results argue that binary break-up and captured star accretion can play a significant role in black hole growth. Of the 91 galaxies from \citet{hannah2024}, the majority experience a growth rate greater than 10\%\ of the SMBH mass per gigayear in one or the other of our models. Eleven galaxies harbor SMBHs that grow as quickly in both of our models. Our analysis of the putative black hole in the Large Magellanic Cloud, \LMCstar, together with direct evidence for binary break-up \citep{lmc}, indicates that roughly one third of its mass grew through this channel. Upcoming observations of HVSs in the LMC, together with the anticipated trove of TDE detections with the Vera C.~Rubin Observatory \citep{bricman2020}, will place meaningful constraints on our predictions.

Our models of black hole growth have limitations. For example, we do not consider the dependence of the fraction of stellar binaries on the stellar mass density or the full phase-space distribution of stars that can interact with an SMBH. We also do not track the bound stars from binary break-up as they accrete onto the black hole. Although we adopt a conservative value for the fraction of binary stars and include a model that compares well with sophisticated numerical calculations of TDE rates \citep{hannah2025}, missing details may affect inferred SMBH growth rates. 

The scenarios described here, with steady-state rates of binary star-SMBH encounters, may be extended to include binary captures during merger events of dwarf galaxies and even major mergers between comparably sized galaxies. Our preliminary explorations indicate that such stochastic events might supply a wealth of new stellar binaries into the central regions of galaxies. This possibility may further enhance the potential of the Hills mechanism not only as a signpost of supermassive black holes, but as a way to grow them.

%\acknowledgments
\section*{Acknowledgments} 

We thank C.~Hannah for generously providing the SMBH data that formed the foundation of this analysis, as well as guidance from A.~Seth. An anonymous referee gave feedback that significantly improved this work. ALJ is grateful for support from the University of Utah's Undergraduate Research Opportunity Program. 

\bibliography{main}

\begin{thebibliography}{}
\expandafter\ifx\csname natexlab\endcsname\relax\def\natexlab#1{#1}\fi
\providecommand{\url}[1]{\href{#1}{#1}}
\providecommand{\dodoi}[1]{doi:~\href{http://doi.org/#1}{\nolinkurl{#1}}}
\providecommand{\doeprint}[1]{\href{http://ascl.net/#1}{\nolinkurl{http://ascl.net/#1}}}
\providecommand{\doarXiv}[1]{\href{https://arxiv.org/abs/#1}{\nolinkurl{https://arxiv.org/abs/#1}}}

\bibitem[{{Antonini} {et~al.}(2010){Antonini}, {Faber}, {Gualandris}, \&
  {Merritt}}]{antonini2010}
{Antonini}, F., {Faber}, J., {Gualandris}, A., \& {Merritt}, D. 2010, \apj,
  713, 90, \dodoi{10.1088/0004-637X/713/1/90}

\bibitem[{{Bahcall} \& {Wolf}(1976)}]{bahcall1976}
{Bahcall}, J.~N., \& {Wolf}, R.~A. 1976, \apj, 209, 214, \dodoi{10.1086/154711}

\bibitem[{{Bernardi} {et~al.}(2003){Bernardi}, {Sheth}, {Annis}, {Burles},
  {Eisenstein}, {Finkbeiner}, {Hogg}, {Lupton}, {Schlegel}, {SubbaRao},
  {Bahcall}, {Blakeslee}, {Brinkmann}, {Castander}, {Connolly}, {Csabai},
  {Doi}, {Fukugita}, {Frieman}, {Heckman}, {Hennessy}, {Ivezi{\'c}}, {Knapp},
  {Lamb}, {McKay}, {Munn}, {Nichol}, {Okamura}, {Schneider}, {Thakar}, \&
  {York}}]{bernardi2003}
{Bernardi}, M., {Sheth}, R.~K., {Annis}, J., {et~al.} 2003, \aj, 125, 1817,
  \dodoi{10.1086/367776}

\bibitem[{{Bondi} \& {Hoyle}(1944)}]{bondi1944}
{Bondi}, H., \& {Hoyle}, F. 1944, \mnras, 104, 273,
  \dodoi{10.1093/mnras/104.5.273}

\bibitem[{{Bortolas} {et~al.}(2023){Bortolas}, {Ryu}, {Broggi}, \&
  {Sesana}}]{bortolas2023}
{Bortolas}, E., {Ryu}, T., {Broggi}, L., \& {Sesana}, A. 2023, \mnras, 524,
  3026, \dodoi{10.1093/mnras/stad2024}

\bibitem[{{Bradnick} {et~al.}(2017){Bradnick}, {Mandel}, \&
  {Levin}}]{bradnick2017}
{Bradnick}, B., {Mandel}, I., \& {Levin}, Y. 2017, \mnras, 469, 2042,
  \dodoi{10.1093/mnras/stx1007}

\bibitem[{{Bricman} \& {Gomboc}(2020)}]{bricman2020}
{Bricman}, K., \& {Gomboc}, A. 2020, \apj, 890, 73,
  \dodoi{10.3847/1538-4357/ab6989}

\bibitem[{{Bromley} {et~al.}(2006){Bromley}, {Kenyon}, {Geller}, {Barcikowski},
  {Brown}, \& {Kurtz}}]{bromley2006}
{Bromley}, B.~C., {Kenyon}, S.~J., {Geller}, M.~J., {et~al.} 2006, \apj, 653,
  1194, \dodoi{10.1086/508419}

\bibitem[{{Bromley} {et~al.}(2012){Bromley}, {Kenyon}, {Geller}, \&
  {Brown}}]{bromley2012}
{Bromley}, B.~C., {Kenyon}, S.~J., {Geller}, M.~J., \& {Brown}, W.~R. 2012,
  \apjl, 749, L42, \dodoi{10.1088/2041-8205/749/2/L42}

\bibitem[{{Brown} {et~al.}(2012){Brown}, {Geller}, \& {Kenyon}}]{brown2012}
{Brown}, W.~R., {Geller}, M.~J., \& {Kenyon}, S.~J. 2012, \apj, 751, 55,
  \dodoi{10.1088/0004-637X/751/1/55}

\bibitem[{{Brown} {et~al.}(2009){Brown}, {Geller}, {Kenyon}, \&
  {Bromley}}]{brown2009}
{Brown}, W.~R., {Geller}, M.~J., {Kenyon}, S.~J., \& {Bromley}, B.~C. 2009,
  \apjl, 690, L69, \dodoi{10.1088/0004-637X/690/1/L69}

\bibitem[{{Brown} {et~al.}(2005){Brown}, {Geller}, {Kenyon}, \&
  {Kurtz}}]{brown2005}
{Brown}, W.~R., {Geller}, M.~J., {Kenyon}, S.~J., \& {Kurtz}, M.~J. 2005,
  \apjl, 622, L33, \dodoi{10.1086/429378}

\bibitem[{{Brown} {et~al.}(2006){Brown}, {Geller}, {Kenyon}, \&
  {Kurtz}}]{brown2006}
---. 2006, \apjl, 640, L35, \dodoi{10.1086/503279}

\bibitem[{{Chen} {et~al.}(2023){Chen}, {Yu}, \& {Lu}}]{correlations2023}
{Chen}, Y., {Yu}, Q., \& {Lu}, Y. 2023, \apj, 952, 135,
  \dodoi{10.3847/1538-4357/acd753}

\bibitem[{{Cohn} \& {Kulsrud}(1978)}]{cohn1978}
{Cohn}, H., \& {Kulsrud}, R.~M. 1978, \apj, 226, 1087, \dodoi{10.1086/156685}

\bibitem[{{Dalton} \& {Sarazin}(1995)}]{dalton1995}
{Dalton}, W.~W., \& {Sarazin}, C.~L. 1995, \apj, 448, 369,
  \dodoi{10.1086/175968}

\bibitem[{{Di Matteo} {et~al.}(2003){Di Matteo}, {Croft}, {Springel}, \&
  {Hernquist}}]{dimatteo2003}
{Di Matteo}, T., {Croft}, R. A.~C., {Springel}, V., \& {Hernquist}, L. 2003,
  \apj, 593, 56, \dodoi{10.1086/376501}

\bibitem[{{Duch{\^e}ne} \& {Kraus}(2013)}]{duchene2013}
{Duch{\^e}ne}, G., \& {Kraus}, A. 2013, \araa, 51, 269,
  \dodoi{10.1146/annurev-astro-081710-102602}

\bibitem[{{Evans} \& {Kochanek}(1989)}]{evans1989}
{Evans}, C.~R., \& {Kochanek}, C.~S. 1989, \apjl, 346, L13,
  \dodoi{10.1086/185567}

\bibitem[{{Event Horizon Telescope Collaboration} {et~al.}(2022){Event Horizon
  Telescope Collaboration}, {Akiyama}, {Alberdi}, {Alef}, {Algaba}, {Anantua},
  {Asada}, {Azulay}, {Bach}, {Baczko}, {Ball}, {Balokovi{\'c}}, {Barrett},
  {Baub{\"o}ck}, {Benson}, {Bintley}, {Blackburn}, {Blundell}, {Bouman},
  {Bower}, {Boyce}, {Bremer}, {Brinkerink}, {Brissenden}, {Britzen},
  {Broderick}, {Broguiere}, {Bronzwaer}, {Bustamante}, {Byun}, {Carlstrom},
  {Ceccobello}, {Chael}, {Chan}, {Chatterjee}, {Chatterjee}, {Chen}, {Chen},
  {Cheng}, {Cho}, {Christian}, {Conroy}, {Conway}, {Cordes}, {Crawford},
  {Crew}, {Cruz-Osorio}, {Cui}, {Davelaar}, {De Laurentis}, {Deane}, {Dempsey},
  {Desvignes}, {Dexter}, {Dhruv}, {Doeleman}, {Dougal}, {Dzib}, {Eatough},
  {Emami}, {Falcke}, {Farah}, {Fish}, {Fomalont}, {Ford}, {Fraga-Encinas},
  {Freeman}, {Friberg}, {Fromm}, {Fuentes}, {Galison}, {Gammie}, {Garc{\'\i}a},
  {Gentaz}, {Georgiev}, {Goddi}, {Gold}, {G{\'o}mez-Ruiz}, {G{\'o}mez}, {Gu},
  {Gurwell}, {Hada}, {Haggard}, {Haworth}, {Hecht}, {Hesper}, {Heumann}, {Ho},
  {Ho}, {Honma}, {Huang}, {Huang}, {Hughes}, {Ikeda}, {Impellizzeri}, {Inoue},
  {Issaoun}, {James}, {Jannuzi}, {Janssen}, {Jeter}, {Jiang},
  {Jim{\'e}nez-Rosales}, {Johnson}, {Jorstad}, {Joshi}, {Jung}, {Karami},
  {Karuppusamy}, {Kawashima}, {Keating}, {Kettenis}, {Kim}, {Kim}, {Kim},
  {Kim}, {Kino}, {Koay}, {Kocherlakota}, {Kofuji}, {Koch}, {Koyama}, {Kramer},
  {Kramer}, {Krichbaum}, {Kuo}, {La Bella}, {Lauer}, {Lee}, {Lee}, {Leung},
  {Levis}, {Li}, {Lico}, {Lindahl}, {Lindqvist}, {Lisakov}, {Liu}, {Liu},
  {Liuzzo}, {Lo}, {Lobanov}, {Loinard}, {Lonsdale}, {Lu}, {Mao}, {Marchili},
  {Markoff}, {Marrone}, {Marscher}, {Mart{\'\i}-Vidal}, {Matsushita},
  {Matthews}, {Medeiros}, {Menten}, {Michalik}, {Mizuno}, {Mizuno}, {Moran},
  {Moriyama}, {Moscibrodzka}, {M{\"u}ller}, {Mus}, {Musoke}, {Myserlis},
  {Nadolski}, {Nagai}, {Nagar}, {Nakamura}, {Narayan}, {Narayanan},
  {Natarajan}, {Nathanail}, {Fuentes}, {Neilsen}, {Neri}, {Ni}, {Noutsos},
  {Nowak}, {Oh}, {Okino}, {Olivares}, {Ortiz-Le{\'o}n}, {Oyama}, {{\"O}zel},
  {Palumbo}, {Paraschos}, {Park}, {Parsons}, {Patel}, {Pen}, {Pesce},
  {Pi{\'e}tu}, {Plambeck}, {PopStefanija}, {Porth}, {P{\"o}tzl}, {Prather},
  {Preciado-L{\'o}pez}, \& {Psaltis}}]{eht2022}
{Event Horizon Telescope Collaboration}, {Akiyama}, K., {Alberdi}, A., {et~al.}
  2022, \apjl, 930, L12, \dodoi{10.3847/2041-8213/ac6674}

\bibitem[{{Faber} \& {Jackson}(1976)}]{faber1976}
{Faber}, S.~M., \& {Jackson}, R.~E. 1976, \apj, 204, 668,
  \dodoi{10.1086/154215}

\bibitem[{{Faber} {et~al.}(1997){Faber}, {Tremaine}, {Ajhar}, {Byun},
  {Dressler}, {Gebhardt}, {Grillmair}, {Kormendy}, {Lauer}, \&
  {Richstone}}]{faber1997}
{Faber}, S.~M., {Tremaine}, S., {Ajhar}, E.~A., {et~al.} 1997, \aj, 114, 1771,
  \dodoi{10.1086/118606}

\bibitem[{{Frank} \& {Rees}(1976)}]{frank1976}
{Frank}, J., \& {Rees}, M.~J. 1976, \mnras, 176, 633,
  \dodoi{10.1093/mnras/176.3.633}

\bibitem[{{Gaia Collaboration} {et~al.}(2023){Gaia Collaboration}, {Vallenari},
  {Brown}, {Prusti}, {de Bruijne}, {Arenou}, {Babusiaux}, {Biermann},
  {Creevey}, {Ducourant}, {Evans}, {Eyer}, {Guerra}, {Hutton}, {Jordi},
  {Klioner}, {Lammers}, {Lindegren}, {Luri}, {Mignard}, {Panem}, {Pourbaix},
  {Randich}, {Sartoretti}, {Soubiran}, {Tanga}, {Walton}, {Bailer-Jones},
  {Bastian}, {Drimmel}, {Jansen}, {Katz}, {Lattanzi}, {van Leeuwen}, {Bakker},
  {Cacciari}, {Casta{\~n}eda}, {De Angeli}, {Fabricius}, {Fouesneau},
  {Fr{\'e}mat}, {Galluccio}, {Guerrier}, {Heiter}, {Masana}, {Messineo},
  {Mowlavi}, {Nicolas}, {Nienartowicz}, {Pailler}, {Panuzzo}, {Riclet}, {Roux},
  {Seabroke}, {Sordo}, {Th{\'e}venin}, {Gracia-Abril}, {Portell}, {Teyssier},
  {Altmann}, {Andrae}, {Audard}, {Bellas-Velidis}, {Benson}, {Berthier},
  {Blomme}, {Burgess}, {Busonero}, {Busso}, {C{\'a}novas}, {Carry}, {Cellino},
  {Cheek}, {Clementini}, {Damerdji}, {Davidson}, {de Teodoro}, {Nu{\~n}ez
  Campos}, {Delchambre}, {Dell'Oro}, {Esquej}, {Fern{\'a}ndez-Hern{\'a}ndez},
  {Fraile}, {Garabato}, {Garc{\'\i}a-Lario}, {Gosset}, {Haigron}, {Halbwachs},
  {Hambly}, {Harrison}, {Hern{\'a}ndez}, {Hestroffer}, {Hodgkin}, {Holl},
  {Jan{\ss}en}, {Jevardat de Fombelle}, {Jordan}, {Krone-Martins}, {Lanzafame},
  {L{\"o}ffler}, {Marchal}, {Marrese}, {Moitinho}, {Muinonen}, {Osborne},
  {Pancino}, {Pauwels}, {Recio-Blanco}, {Reyl{\'e}}, {Riello}, {Rimoldini},
  {Roegiers}, {Rybizki}, {Sarro}, {Siopis}, {Smith}, {Sozzetti}, {Utrilla},
  {van Leeuwen}, {Abbas}, {{\'A}brah{\'a}m}, {Abreu Aramburu}, {Aerts},
  {Aguado}, {Ajaj}, {Aldea-Montero}, {Altavilla}, {{\'A}lvarez}, {Alves},
  {Anders}, {Anderson}, {Anglada Varela}, {Antoja}, {Baines}, {Baker},
  {Balaguer-N{\'u}{\~n}ez}, {Balbinot}, {Balog}, {Barache}, {Barbato},
  {Barros}, {Barstow}, {Bartolom{\'e}}, {Bassilana}, {Bauchet}, {Becciani},
  {Bellazzini}, {Berihuete}, {Bernet}, {Bertone}, {Bianchi}, {Binnenfeld},
  {Blanco-Cuaresma}, {Blazere}, {Boch}, {Bombrun}, {Bossini}, {Bouquillon},
  {Bragaglia}, {Bramante}, {Breedt}, {Bressan}, {Brouillet}, {Brugaletta},
  {Bucciarelli}, {Burlacu}, {Butkevich}, {Buzzi}, {Caffau}, {Cancelliere},
  {Cantat-Gaudin}, {Carballo}, {Carlucci}, {Carnerero}, {Carrasco},
  {Casamiquela}, {Castellani}, {Castro-Ginard}, {Chaoul}, {Charlot}, {Chemin},
  {Chiaramida}, {Chiavassa}, {Chornay}, {Comoretto}, {Contursi}, {Cooper},
  {Cornez}, {Cowell}, {Crifo}, {Cropper}, {Crosta}, {Crowley}, {Dafonte},
  {Dapergolas}, {David}, {David}, {de Laverny}, {De Luise}, \& {De
  March}}]{gaiadr32023}
{Gaia Collaboration}, {Vallenari}, A., {Brown}, A.~G.~A., {et~al.} 2023, \aap,
  674, A1, \dodoi{10.1051/0004-6361/202243940}

\bibitem[{{Genzel} {et~al.}(2010){Genzel}, {Eisenhauer}, \&
  {Gillessen}}]{genzel2010}
{Genzel}, R., {Eisenhauer}, F., \& {Gillessen}, S. 2010, Reviews of Modern
  Physics, 82, 3121, \dodoi{10.1103/RevModPhys.82.3121}

\bibitem[{{Gezari}(2021)}]{gezari2021}
{Gezari}, S. 2021, \araa, 59, 21, \dodoi{10.1146/annurev-astro-111720-030029}

\bibitem[{{Ghez} {et~al.}(2008){Ghez}, {Salim}, {Weinberg}, {Lu}, {Do}, {Dunn},
  {Matthews}, {Morris}, {Yelda}, {Becklin}, {Kremenek}, {Milosavljevic}, \&
  {Naiman}}]{ghez2008}
{Ghez}, A.~M., {Salim}, S., {Weinberg}, N.~N., {et~al.} 2008, \apj, 689, 1044,
  \dodoi{10.1086/592738}

\bibitem[{{Gonz{\'a}lez Prieto} {et~al.}(2025){Gonz{\'a}lez Prieto},
  {Rodriguez}, \& {Cabrera}}]{gonzalezprieto2025}
{Gonz{\'a}lez Prieto}, E., {Rodriguez}, C.~L., \& {Cabrera}, T. 2025, arXiv
  e-prints, arXiv:2507.06316, \dodoi{10.48550/arXiv.2507.06316}

\bibitem[{{Gould} \& {Quillen}(2003)}]{gould2003}
{Gould}, A., \& {Quillen}, A.~C. 2003, \apj, 592, 935, \dodoi{10.1086/375840}

\bibitem[{{Greene} {et~al.}(2020){Greene}, {Strader}, \& {Ho}}]{greene2020}
{Greene}, J.~E., {Strader}, J., \& {Ho}, L.~C. 2020, \araa, 58, 257,
  \dodoi{10.1146/annurev-astro-032620-021835}

\bibitem[{{G{\"u}ltekin} {et~al.}(2009){G{\"u}ltekin}, {Richstone}, {Gebhardt},
  {Lauer}, {Pinkney}, {Aller}, {Bender}, {Dressler}, {Faber}, {Filippenko},
  {Green}, {Ho}, {Kormendy}, \& {Siopis}}]{gultekin2009}
{G{\"u}ltekin}, K., {Richstone}, D.~O., {Gebhardt}, K., {et~al.} 2009, \apj,
  695, 1577, \dodoi{10.1088/0004-637X/695/2/1577}

\bibitem[{{Han} {et~al.}(2025){Han}, {El-Badry}, {Lucchini}, {Hernquist},
  {Brown}, {Garavito-Camargo}, {Conroy}, \& {Sari}}]{lmc}
{Han}, J.~J., {El-Badry}, K., {Lucchini}, S., {et~al.} 2025, arXiv e-prints,
  arXiv:2502.00102, \dodoi{10.48550/arXiv.2502.00102}

\bibitem[{{Hannah}(2025)}]{reptide}
{Hannah}, C.~H. 2025, REPTiDE (Zenodo), \dodoi{10.5281/zenodo.14969335}

\bibitem[{{Hannah} {et~al.}(2024){Hannah}, {Seth}, {Stone}, \& {van
  Velzen}}]{hannah2024}
{Hannah}, C.~H., {Seth}, A.~C., {Stone}, N.~C., \& {van Velzen}, S. 2024, \aj,
  168, 137, \dodoi{10.3847/1538-3881/ad630a}

\bibitem[{{Hannah} {et~al.}(2025){Hannah}, {Stone}, {Seth}, \& {van
  Velzen}}]{hannah2025}
{Hannah}, C.~H., {Stone}, N.~C., {Seth}, A.~C., \& {van Velzen}, S. 2025, \apj,
  988, 29, \dodoi{10.3847/1538-4357/addd1b}

\bibitem[{{Hills}(1975)}]{hills1975}
{Hills}, J.~G. 1975, \nat, 254, 295, \dodoi{10.1038/254295a0}

\bibitem[{{Hills}(1988)}]{hills1988}
---. 1988, \nat, 331, 687, \dodoi{10.1038/331687a0}

\bibitem[{{Jin} {et~al.}(2025){Jin}, {Pasquato}, {Davis}, {Deleu}, {Luo},
  {Cho}, {Lemos}, {Perreault-Levasseur}, {Bengio}, {Kang}, {Macci{\`o}}, \&
  {Hezaveh}}]{jin2025}
{Jin}, Z., {Pasquato}, M., {Davis}, B.~L., {et~al.} 2025, \apj, 979, 212,
  \dodoi{10.3847/1538-4357/ad9ded}

\bibitem[{{Kormendy} \& {Ho}(2013)}]{kormendy2013}
{Kormendy}, J., \& {Ho}, L.~C. 2013, \araa, 51, 511,
  \dodoi{10.1146/annurev-astro-082708-101811}

\bibitem[{{Lightman} \& {Shapiro}(1977)}]{lightman1977}
{Lightman}, A.~P., \& {Shapiro}, S.~L. 1977, \apj, 211, 244,
  \dodoi{10.1086/154925}

\bibitem[{{Magorrian} \& {Tremaine}(1999)}]{magorrian1999}
{Magorrian}, J., \& {Tremaine}, S. 1999, \mnras, 309, 447,
  \dodoi{10.1046/j.1365-8711.1999.02853.x}

\bibitem[{{McConnell} \& {Ma}(2013)}]{mcconnel2013}
{McConnell}, N.~J., \& {Ma}, C.-P. 2013, \apj, 764, 184,
  \dodoi{10.1088/0004-637X/764/2/184}

\bibitem[{{Merritt}(2013)}]{merritt2013}
{Merritt}, D. 2013, {Dynamics and Evolution of Galactic Nuclei} (Princeton:
  {Princeton University Press})

\bibitem[{{Milosavljevi{\'c}} {et~al.}(2006){Milosavljevi{\'c}}, {Merritt}, \&
  {Ho}}]{milosavljevi2006}
{Milosavljevi{\'c}}, M., {Merritt}, D., \& {Ho}, L.~C. 2006, \apj, 652, 120,
  \dodoi{10.1086/508134}

\bibitem[{{Moe} \& {Di Stefano}(2017)}]{moe2017}
{Moe}, M., \& {Di Stefano}, R. 2017, \apjs, 230, 15,
  \dodoi{10.3847/1538-4365/aa6fb6}

\bibitem[{{Montesinos Armijo} \& {de Freitas
  Pacheco}(2011)}]{montesinosarmijo2011}
{Montesinos Armijo}, M.~A., \& {de Freitas Pacheco}, J.~A. 2011, \aap, 526,
  A146, \dodoi{10.1051/0004-6361/201015026}

\bibitem[{{Olejak} {et~al.}(2025){Olejak}, {Stegmann}, {de Mink}, {Valli},
  {Sari}, {Justham}, \& {Ryu}}]{olejak2025}
{Olejak}, A., {Stegmann}, J., {de Mink}, S.~E., {et~al.} 2025, \apjl, 987, L11,
  \dodoi{10.3847/2041-8213/ade432}

\bibitem[{{Perets} {et~al.}(2007){Perets}, {Hopman}, \&
  {Alexander}}]{perets2007}
{Perets}, H.~B., {Hopman}, C., \& {Alexander}, T. 2007, \apj, 656, 709,
  \dodoi{10.1086/510377}

\bibitem[{{Rees}(1988)}]{rees1988}
{Rees}, M.~J. 1988, \nat, 333, 523, \dodoi{10.1038/333523a0}

\bibitem[{{Rossi} {et~al.}(2014){Rossi}, {Kobayashi}, \& {Sari}}]{rossi2014}
{Rossi}, E.~M., {Kobayashi}, S., \& {Sari}, R. 2014, \apj, 795, 125,
  \dodoi{10.1088/0004-637X/795/2/125}

\bibitem[{{Ryu} {et~al.}(2022){Ryu}, {Trani}, \& {Leigh}}]{compactsmbhb2022}
{Ryu}, T., {Trani}, A.~A., \& {Leigh}, N. W.~C. 2022, \mnras, 515, 2430,
  \dodoi{10.1093/mnras/stac1987}

\bibitem[{{Sari} {et~al.}(2010){Sari}, {Kobayashi}, \& {Rossi}}]{sari2010}
{Sari}, R., {Kobayashi}, S., \& {Rossi}, E. 2010, \apj, 708, 605,
  \dodoi{10.1088/0004-637X/708/1/605}

\bibitem[{{Sellwood} \& {Wilkinson}(1993)}]{sellwood1993}
{Sellwood}, J.~A., \& {Wilkinson}, A. 1993, Reports on Progress in Physics, 56,
  173, \dodoi{10.1088/0034-4885/56/2/001}

\bibitem[{{Sersante} {et~al.}(2025){Sersante}, {Penoyre}, \&
  {Rossi}}]{sersante2025}
{Sersante}, B., {Penoyre}, Z., \& {Rossi}, E.~M. 2025, \mnras, 544, 1688,
  \dodoi{10.1093/mnras/staf1766}

\bibitem[{{Silk} \& {Rees}(1998)}]{silk1998}
{Silk}, J., \& {Rees}, M.~J. 1998, \aap, 331, L1,
  \dodoi{10.48550/arXiv.astro-ph/9801013}

\bibitem[{{Stone} {et~al.}(2017){Stone}, {K{\"u}pper}, \&
  {Ostriker}}]{stone2017}
{Stone}, N.~C., {K{\"u}pper}, A. H.~W., \& {Ostriker}, J.~P. 2017, \mnras, 467,
  4180, \dodoi{10.1093/mnras/stx097}

\bibitem[{{Stone} \& {Metzger}(2016)}]{stone2016}
{Stone}, N.~C., \& {Metzger}, B.~D. 2016, \mnras, 455, 859,
  \dodoi{10.1093/mnras/stv2281}

\bibitem[{{Thomsen} {et~al.}(2022){Thomsen}, {Kwan}, {Dai}, {Wu}, {Roth}, \&
  {Ramirez-Ruiz}}]{thomsen2022}
{Thomsen}, L.~L., {Kwan}, T.~M., {Dai}, L., {et~al.} 2022, \apjl, 937, L28,
  \dodoi{10.3847/2041-8213/ac911f}

\bibitem[{{Toscani} {et~al.}(2025){Toscani}, {Broggi}, {Sesana}, \&
  {Rossi}}]{toscani2025}
{Toscani}, M., {Broggi}, L., {Sesana}, A., \& {Rossi}, E.~M. 2025, \aap, 703,
  A75, \dodoi{10.1051/0004-6361/202555648}

\bibitem[{{Vasiliev}(2018)}]{vasiliev2018}
{Vasiliev}, E. 2018, \mnras, 481, L100, \dodoi{10.1093/mnrasl/sly168}

\bibitem[{{Verberne} {et~al.}(2025){Verberne}, {Koposov}, {Rossi}, \&
  {Penoyre}}]{verberne2025}
{Verberne}, S., {Koposov}, S.~E., {Rossi}, E.~M., \& {Penoyre}, Z. 2025, arXiv
  e-prints, arXiv:2506.19570, \dodoi{10.48550/arXiv.2506.19570}

\bibitem[{{Webb} \& {Foustoul}(2022)}]{observation2022}
{Webb}, N.~A., \& {Foustoul}, V. 2022, in SF2A-2022: Proceedings of the Annual
  meeting of the French Society of Astronomy and Astrophysics. Eds.: J.
  Richard, ed. J.~{Richard}, A.~{Siebert}, E.~{Lagadec}, N.~{Lagarde},
  O.~{Venot}, J.~{Malzac}, J.~B. {Marquette}, M.~{N'Diaye}, \& B.~{Briot},
  251--254

\bibitem[{{Yao} {et~al.}(2023){Yao}, {Ravi}, {Gezari}, {van Velzen}, {Lu},
  {Schulze}, {Somalwar}, {Kulkarni}, {Hammerstein}, {Nicholl}, {Graham},
  {Perley}, {Cenko}, {Stein}, {Ricarte}, {Chadayammuri}, {Quataert}, {Bellm},
  {Bloom}, {Dekany}, {Drake}, {Groom}, {Mahabal}, {Prince}, {Riddle},
  {Rusholme}, {Sharma}, {Sollerman}, \& {Yan}}]{yao2023}
{Yao}, Y., {Ravi}, V., {Gezari}, S., {et~al.} 2023, \apjl, 955, L6,
  \dodoi{10.3847/2041-8213/acf216}

\bibitem[{{Yu} \& {Lai}(2024)}]{yu2024}
{Yu}, F., \& {Lai}, D. 2024, \apj, 977, 268, \dodoi{10.3847/1538-4357/ad93a6}

\bibitem[{{Yu} \& {Tremaine}(2002)}]{yu2002}
{Yu}, Q., \& {Tremaine}, S. 2002, \mnras, 335, 965,
  \dodoi{10.1046/j.1365-8711.2002.05532.x}

\bibitem[{{Yu} \& {Tremaine}(2003)}]{yu2003}
---. 2003, \apj, 599, 1129, \dodoi{10.1086/379546}

\bibitem[{{Zheng} {et~al.}(2014){Zheng}, {Carlin}, {Beers}, {Deng},
  {Grillmair}, {Guhathakurta}, {L{\'e}pine}, {Newberg}, {Yanny}, {Zhang},
  {Liu}, {Jin}, \& {Zhang}}]{zheng2014}
{Zheng}, Z., {Carlin}, J.~L., {Beers}, T.~C., {et~al.} 2014, \apjl, 785, L23,
  \dodoi{10.1088/2041-8205/785/2/L23}

\end{thebibliography}
\bibliographystyle{aasjournal}

\end{document}